\documentclass[conference]{IEEEtran}
\IEEEoverridecommandlockouts
\usepackage{cite}
\usepackage{amsmath,amssymb,amsfonts}
\usepackage{algorithmic}
\usepackage{graphicx}
\usepackage{textcomp}
\usepackage{xcolor}
\def\BibTeX{{\rm B\kern-.05em{\sc i\kern-.025em b}\kern-.08em
    T\kern-.1667em\lower.7ex\hbox{E}\kern-.125emX}}

\usepackage{todonotes}
\usepackage{multirow}
\usepackage{caption}
\usepackage{subcaption}
\usepackage{colortbl}
\usepackage{graphicx}
\usepackage{algorithm}
\usepackage[hidelinks]{hyperref}
\usepackage[multiple]{footmisc}

\let\oldFootnote\footnote
\newcommand\nextToken\relax

\renewcommand\footnote[1]{%
    \oldFootnote{#1}\futurelet\nextToken\isFootnote}

\newcommand\isFootnote{%
    \ifx\footnote\nextToken\textsuperscript{,}\fi}

\floatname{algorithm}{Example}

\begin{document}

\title{A Preliminary Study on a Conceptual Game Feature Generation 
and Recommendation System}


\makeatletter
\newcommand{\linebreakand}{%
  \end{@IEEEauthorhalign}
  \hfill\mbox{}\par
  \mbox{}\hfill\begin{@IEEEauthorhalign}
}
\makeatother

\author{\IEEEauthorblockN{M Charity}
\IEEEauthorblockA{\textit{Game Innovation Lab} \\
\textit{New York Univesity}\\
Brooklyn, US \\
mlc761@nyu.edu}
\and
\IEEEauthorblockN{Yash Bhartia}
\IEEEauthorblockA{\textit{BITS Pilani}\\
\textit{K.K. Birla Goa Campus}\\
Goa, India \\
f20190151@goa.bits-pilani.ac.in}
\and
\IEEEauthorblockN{Daniel Zhang}
\IEEEauthorblockA{\textit{Game Innovation Lab} \\
\textit{New York Univesity}\\
Brooklyn, US \\
dwz206@nyu.edu}
\linebreakand 
\IEEEauthorblockN{Ahmed Khalifa}
\IEEEauthorblockA{\textit{Institute of Digital Games} \\
\textit{University of Malta}\\
Gzira, Malta \\
ahmed.khalifa@um.edu.mt}
\and
\IEEEauthorblockN{Julian Togelius}
\IEEEauthorblockA{\textit{Game Innovation Lab} \\
\textit{New York Univesity}\\
Brooklyn, US \\
julian@togelius.com}
}


\IEEEoverridecommandlockouts

\IEEEpubid{\makebox[\columnwidth]{979-8-3503-2277-4/23/\$31.00~\copyright2023 IEEE \hfill} \hspace{\columnsep}\makebox[\columnwidth]{ }}

\maketitle

\IEEEpubidadjcol

\begin{abstract}
This paper introduces a system used to generate game feature suggestions based on a text prompt. Trained on the game descriptions of almost 60k games, it uses the word embeddings of a small GLoVe model to extract features and entities found in thematically similar games which are then passed through a generator model to generate new features for a user's prompt. We perform a short user study comparing the features generated from a fine-tuned GPT-2 model, a model using the ConceptNet, and human-authored game features. Although human suggestions won the overall majority of votes, the GPT-2 model outperformed the human suggestions in certain games. This system is part of a larger game design assistant tool that is able to collaborate with users at a conceptual level.
\end{abstract}


\section{Introduction}
The game design process is an expressive, creative, and challenging task for artists and developers. Most games start from a base idea focused on a game mechanic, a visual style, or a general theme. These ideas can come from anywhere, including other games, art mediums, and personal experiences. Many independent game designers start by developing a prototype (for game jams or as a hobby~\cite{diver2016indie}) before committing to developing the full game in a short time span. 
While developing a prototype, designers sometimes seek feedback and suggestions for new game features from their team, friends, or the community, in order to find exciting new features to enrich the main game idea~\cite{lassheikki2019game}. As such, these games can take on more personal qualities as the people who helped add a piece of their own experience and creativity to it. These range from small, abstract ways such as suggesting or voting on game features they find fun or interesting (e.g. ``Minecraft" mob votes\footnote{\url{https://minecraft.fandom.com/wiki/Mob_Vote}}), or in a way that more directly impacts the final product such as creating game elements\footnote{\url{https://distractionware.com/blog/2022/07/the-reunion-reunion/}}. 

In this paper, we introduce an AI system that simulates the role of friends and communities generating game features from user-provided text prompts describing their game. This subsystem provides a novel form of abstract content generation that ideally would aid the game designer during the design process. We performed an informal user study to evaluate the output of 2 different generative models - a fine-tuned GPT-2 model and a tree search-based model using ConceptNet - against human-authored game feature suggestions to examine which suggestions users prefer aesthetically given a variety of both generic and niche themed game descriptions.

\section{Background}
We use Mullich's definition of ``game features" which are the descriptive aspects of a game’s design, art, or technical capabilities to explain the overall game experience~\cite{mullich_2018}. These features are typically found in the game's summary text and use higher-level language to communicate the most interesting aspects of the game. ``Game features" is more broadly compared to Sicart's definition~\cite{sicart2008defining} for ``game rules'' and ``game mechanics''. Most of the previous work focused on generating game rules/mechanics either autonomously~\cite{cook2013mechanic,khalifa2017general} or in a mixed-initiative fashion~\cite{nelson2017mixed, machado2019pitako}. There is a lack of work that looks to design games from more abstract game features - or works that focus on game design in its early concept stage. Having a system that can work collaboratively with a user during this phase of game design could be beneficial and create a new path of research for mixed-initiative game design.

One notably interesting  direction of work is the use of semantic networks to generate new games. Semantic networks~\cite{sowa1992semantic} represent information as a network, connecting concepts and entities to generate games. Nelson and Mateas ~\cite{nelson2007towards} proposed a system for generating microgames~\cite{gingold2005warioware} using semantic networks by creating a playable game from a text prompt. However, this system was too restrictive and couldn't allow for new innovative game features and ideas.

Recent developments with large language models (LLMs), such as OpenAI's GPT-2~\cite{radford2019better}, allow artists and academics to use these systems for more than text generation. Fine-tuning and prompt-engineering these large language models on customized datasets can guide the models towards structured or thematically relevant outputs~\cite{reynolds2021prompt}. LLMs have been used to generate game content such as narrative quest designs~\cite{vartinen2022generating} and game environments~\cite{frans2021ai}. 

\begin{figure*}[t]
    \centering
    \includegraphics[width=0.8\linewidth]{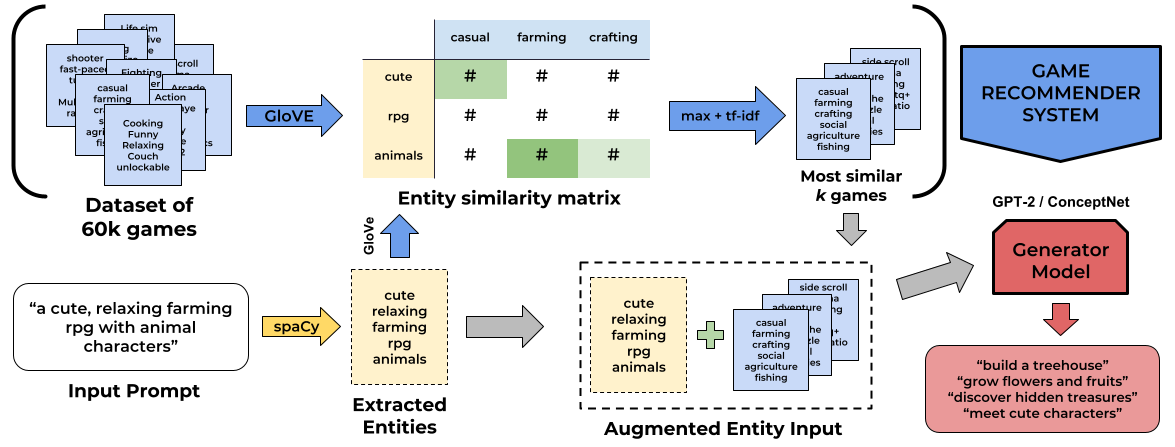}
    \caption{Diagram of the whole system consists of two main components: the recommendation system and the generator model. 
    }
    \label{fig:sys_diagram}
\end{figure*}


\section{System Design}

Our system is made up of 2 parts - a recommendation system for identifying which games are most thematically similar and a generator model that suggests new features (shown in figure~\ref{fig:sys_diagram}). 
For the generator model, we used 2 different models - a fine-tuned GPT-2 LLM and the semantic network API ConceptNet. The following subsections describe these systems and the training data used to fine-tune them.


\subsection{Training Data}
\label{traindat}

Using the Steam Web API\footnote{https://steamcommunity.com/dev/apiterms} and LaunchBox Games Database\footnote{http://gamesdb.launchbox-app.com/Metadata.zip}, we were able to extract tags and summary descriptions for over 400k published games. The SpaCy\footnote{\url{https://spacy.io/}} library's part-of-speech tokenizer was used to extract nouns from the description text to use as theme-relevant entities. As Mullich stated, game features are often found in the game's descriptive summary text, typically starting with a verb (i.e. `jump', `build', or `attack') and ending with a noun (i.e. `platform', `tower', `enemy'.) Therefore, we searched this information using part-of-speech grammar expressions to extract ``feature"-like phrases (VERB / ARTICLE? / NOUN). Games that did not contain any features found using the grammar were removed from the training set. This left us with nearly 60k games that contained a list of entity-based theme-relevant nouns, pre-defined labeled genre tags, and grammar-extracted feature phrases to be used as the training data.

\subsection{Recommendation System}\label{recsys}
An input prompt in this system consists of only a short sentence describing high-level ideas of the user's game. However, there is not nearly enough information to match it to the games in the training data. Therefore, we created a game recommendation system to find a set of games that are closest in theme and gameplay to the user's game and then augment the prompt with more information.

We used SpaCy to tokenize the input prompt in the same fashion as the training data. After that, the recommendation system uses the GloVe word embedding model\cite{pennington2014glove} to encode each noun entity in the input prompt and the training data. This gives us a 50 dimension embedding vector for every noun entity. We then use these vectors to calculate a similarity matrix between the input prompt vectors and the vectors of each game in the training data using cosine similarity. To calculate the final score for each game, we compute the maximum similarity score between the input vectors and each game vector. These scores are added together using a weighted sum such that the weights for the input nouns are calculated using the term frequency inverse-document frequency (TF-IDF)~\cite{ramos2003using}. This is intended to exploit the uniqueness of particular words that may be seen less often in the game descriptions (i.e. "shooter" would have a lower TF-IDF score compared to ``alligator''.) Finally, the tags and noun entities from this subset of recommended games are fed to the generator model.

\begin{table*}[t]
\centering
\resizebox{0.88\textwidth}{!}{%
\begin{tabular}{|c|l|l|l|l|}
\hline
\rowcolor[HTML]{d1d1d1} 
{\color[HTML]{343434} \begin{tabular}[c]{@{}c@{}}Game \\ \#\end{tabular}} & \multicolumn{1}{c|}{\cellcolor[HTML]{d1d1d1}{\color[HTML]{343434} Prompt}} & \multicolumn{1}{c|}{\cellcolor[HTML]{d1d1d1}{\color[HTML]{343434} Human Features}} & \multicolumn{1}{c|}{\cellcolor[HTML]{d1d1d1}{\color[HTML]{343434} GPT-2 Features}} & \multicolumn{1}{c|}{\cellcolor[HTML]{d1d1d1}{\color[HTML]{343434} ConceptNet Features}} \\ \hline
1 & \textit{\begin{tabular}[c]{@{}l@{}}a class based multiplayer online shooter with \\ capture the flag, death match, and deliver the payload\end{tabular}} & \begin{tabular}[c]{@{}l@{}}shoot gun to heal allies\\ double jump in the air\\ drink energy drinks\\ throw suspicious jars\\ backstab enemy targets\end{tabular} & \begin{tabular}[c]{@{}l@{}}use a bomb\\ help them survive the assault\\ earn experience\\ change the route\\ use an offline password\end{tabular} & \begin{tabular}[c]{@{}l@{}}secure doors \\ pick a leader \\ construct buildings\\ blow something up \\ kill someone\end{tabular} \\ \hline
\rowcolor[HTML]{DAE8FC} 
2 & \textit{\begin{tabular}[c]{@{}l@{}}a 5v5 game where you protect your base while \\ destroying enemy bases using an array of different \\ abilities and items\end{tabular}} & \begin{tabular}[c]{@{}l@{}}attack enemy turrets\\ buy power ups\\ type to communicate with enemy team\\ coordinate tactics with allies\\ protect allied mobs\end{tabular} & \begin{tabular}[c]{@{}l@{}}fight against each other in an arena\\ advance through the game\\ control enemies\\ start with a single character\\ become the champion\end{tabular} & \begin{tabular}[c]{@{}l@{}}secure valuables\\ hold a large conference\\ find the treasure\\ pick a leader\\ plan a road trip\end{tabular} \\ \hline
3 & \textit{\begin{tabular}[c]{@{}l@{}}an open-world exploration game in a post-apocalyptic \\ fantasy world where you can climb anything and \\ destroy everything\end{tabular}} & \begin{tabular}[c]{@{}l@{}}explore the wilds\\ battle towering enemies\\ gather ingredients\\ bundle up with warmer clothes\\ think quickly and develop the right strategies\end{tabular} & \begin{tabular}[c]{@{}l@{}}Travel to a variety of forests and field\\ make weapons\\ learn new combat\\ become the new king\\ create new outfits\end{tabular} & \begin{tabular}[c]{@{}l@{}}shred into pieces of confetti\\ provide comfort\\ prevent starvation\\ house a king\\ find shells on a shoreline sometimes\end{tabular} \\ \hline
\rowcolor[HTML]{DAE8FC} 
4 & \textit{\begin{tabular}[c]{@{}l@{}}an RPG about a princess who collects swords and \\ flowers to turn into potions and is secretly a frog\end{tabular}} & \begin{tabular}[c]{@{}l@{}}meet new animal friends\\ explore the ruined castle\\ identify new flora and fauna\\ discover new alchemy recipes\\ find the antidote to your curse\end{tabular} & \begin{tabular}[c]{@{}l@{}}grow your own plants\\ make your own unique potions\\ sell your creations\\ explore a huge world\\ use your imagination\end{tabular} & \begin{tabular}[c]{@{}l@{}}buy and sell things\\ discover the world\\ make wood pulp\\ reach towards the sky\\ make foods\end{tabular} \\ \hline
5 & \textit{\begin{tabular}[c]{@{}l@{}}a collaborative cooking game where you make \\ and sell onigiri in your college dorm room\end{tabular}} & \begin{tabular}[c]{@{}l@{}}make onigiri on the weekends\\ pay off your tuition\\ decorate your dorm room\\ buy new meats and veggies\\ hire friends and roommates part-time\end{tabular} & \begin{tabular}[c]{@{}l@{}}Sharpen your knives\\ communicate and coordinate NPC actions\\ engage in hard boiled head\\ become a master chef\\ battle your crew\end{tabular} & \begin{tabular}[c]{@{}l@{}}cook a gourmet meal\\ cut things\\ cure a headache\\ cry when cutting an onion\\ furnish your home\end{tabular} \\ \hline
\rowcolor[HTML]{DAE8FC} 
6 & \textit{\begin{tabular}[c]{@{}l@{}}a retro-futuristic cyberpunk skateboarding game where \\ you hack corporations, the robot police, and the \\ street gangs\end{tabular}} & \begin{tabular}[c]{@{}l@{}}hack the corpo suits\\ vandalize properties with graffiti\\ find secret hideouts\\ skate the streets of the city\\ upgrade your bioware tech\end{tabular} & \begin{tabular}[c]{@{}l@{}}experience the iconic grind\\ master the deep combo system\\ choose their personality\\ complete secret missions\\ Show off your style\end{tabular} & \begin{tabular}[c]{@{}l@{}}win a war\\ scare people\\ play science fiction games\\ organize data\\ give money to each other\end{tabular} \\ \hline
\end{tabular}%
}
\caption{Table of hand-authored and model-generated features for each game prompt.}
\label{tab:prompt_feat_set}
\end{table*}

\subsection{GPT-2 Generator Model}
\label{sec:gpt2model}

For the first generator, we fine-tuned a medium GPT-2 model~\cite{radford2019language} using the tag set, extracted entities, and description features of the training dataset. The training was restricted to 5 epochs due to computational constraints and to avoid overfitting. After some preliminary experiments with hyperparameter values and informal evaluation of the generated output, we decided to use the following hyperparameter values for the model during generation: $temperature = 0.95$, $top\_k = 100$, $top\_p = 0.8$, and $repetition\_penalty = 0.95$. 

For new feature generation, the output from the recommendation system is preprocessed and passed to the model (a list of tags and entity nouns). The generator often produced a large list of features - occasionally repeating some features. We manually selected 5 features for each prompt from the output of the model to be used in the user study. This manual selection emulates how a user might pick and choose ideal game features that fit the input prompt better than others.

\subsection{ConceptNet Generator Model}
\label{sec:cn_model}

The second generator model is a search-based semantic model that uses ConceptNet~\cite{speer2017conceptnet} to generate features. The ConceptNet model uses the list of noun entities from a prompt to request the semantic information of each word from the ConceptNet API. Unlike the GPT-2 model, this model directly uses the ConceptNet API to generate features with the need for training. 

To find the features from the ConceptNet data, we find all the semantic connections for our input nouns that are labeled with \textit{``CapableOf''} and \textit{``UsedFor''}. We use the output directly from these lists as they have a similar grammar format to the training data. Similarly, we manually selected five generated features for the user study. For this study, none of the additional game tags suggested by the recommendation system were used as they exponentially increase the execution time, and the output number of generated features --- based on the user prompt --- was more than enough.

\section{Experiment Setup}

Six prompts were created for the user study experiment. The first half of the prompts described existing games with common genres and themes found in triple-A titles: ``Team Fortress 2" - an FPS multiplayer game, ``DOTA 2" - a MOBA game and ``The Legend of Zelda: Breath of the Wild" - a single-player open-world action game. The second half of the prompts described games that have not been made (yet) and with less commonly found genres and more niche themes: a role-playing game with crafting mechanics and magic systems, an economy-based cooking game set in a college dorm room, and a cyberpunk skateboarding game. 
For all 6 games, we hand-authored 5 game features as a third comparison group. Naturally, the fake game prompts were hand-authored since they do not exist. However, despite human-authored features existing for the real games, the GPT-2 model was fine-tuned on these same features and was therefore likely to generate the same features when prompted. As such, we created a new set of features for each of the real game prompts. Table \ref{tab:prompt_feat_set} shows the set of features selected for the experiment for each prompt.

For this small user study, participants voted on which set of features of the 3 groups (human, GPT-2, and ConceptNet) they preferred when provided the one-sentence game description prompts. For each prompt, the 3 sets were anonymized and randomly given the label A, B, or C. A Google Form poll was created for each game prompt asking participants 'Which set of game features do you prefer?.' The authorship of the features was not disclosed to participants and they were not told whether the features they were shown were artificially generated. Participants were also given the option of submitting their own suggestions for new features. The polls were circulated both online via Twitter and Discord and physically as small printout flyers in the lobby of a classroom floor for game design students. The number of votes for each set of features was the focus of the study, therefore no personal information was collected and each response was anonymous.

\section{Results and Discussion}

\begin{table*}[ht!]
    \fontsize{6pt}{6pt}\selectfont
      \centering
        \begin{tabular}{|c|l||c|l|}
        \hline
        \rowcolor[HTML]{d1d1d1} 
        \begin{tabular}[c]{@{}c@{}}Game \\ \#\end{tabular} &
          ChatGPT response (real games) & \begin{tabular}[c]{@{}c@{}}Game \\ \#\end{tabular} &
          ChatGPT response (fake games) \\ \hline
        1 &
          \begin{tabular}[c]{@{}l@{}}Specialized classes with unique abilities\\ Fast-paced gameplay with varied objectives\\ Intense team-based combat\\ Customizable weapons and loadouts\\ Competitive matchmaking and ranking system\end{tabular} & 4 &
  \begin{tabular}[c]{@{}l@{}}Sword collecting and upgrading system\\ Alchemy-based potion crafting mechanics\\ Frog transformation with unique abilities\\ Immersive fantasy world and characters\\ Non-linear story with multiple endings\end{tabular} \\ \hline
        \rowcolor[HTML]{DAE8FC} 
        2 &
          \begin{tabular}[c]{@{}l@{}}Dynamic hero abilities and item builds\\ Base building and defense mechanics\\ Resource management and strategy\\ Intense team-based battles\\ Competitive ranking and matchmaking system\end{tabular} & 5 &
  \begin{tabular}[c]{@{}l@{}}Cooperative gameplay with friends\\ Customizable dorm room and kitchen\\ Onigiri recipe book with unique ingredients\\ Customer feedback and rating system\\ Time and resource management mechanics\end{tabular} \\ \hline
        3 &
          \begin{tabular}[c]{@{}l@{}}Dynamic environmental destruction and interaction\\ Fantasy creatures and unique enemies\\ Character customization and progression\\ Immersive exploration and discovery\\ Non-linear story and quest system\end{tabular} & 6 &
  \begin{tabular}[c]{@{}l@{}}Dynamic skateboarding mechanics and stunts\\ Hacking mini-games and puzzles\\ Open-world cyberpunk city to explore\\ Factions and reputation system\\ Story-driven campaign with multiple endings\end{tabular} \\
          \hline
          \end{tabular}
    \caption{Table of ChatGPT responses given the same prompts used in the user study}
    \label{tab:ChatGPT_out}
\end{table*}



The polling period lasted for 3 days. We collected over 140 responses across the 6 polls - with an average of 23.6 responses per poll. 
The human-authored feature sets had a total majority of votes across all 6 polls with 52\% of total votes. For the real games prompts, the number of votes is nearly equal for the transformer and the human-authored features; with the GPT-2 transformer model's features having a slight majority of votes (28) over the human-authored features (27) across the 3 games. Conversely, with the fake games, the human features had the strong majority of votes (47) over the transformer (15.) The ConceptNet model underperformed in polls except for prompt \#1 - outvoting the GPT-2 model.

This stark difference in votes could be due to the frequency of common game genres in the training data. For example, Game \#2 GPT-2 generated features are similar to those found in other open-world fantasy games, such as 'Elden Ring'. 
On the other hand, the fake prompts were more niche in theme and did not exist conceptually in the training dataset, therefore the artificial models struggled to generate features that were related to these genres and themes. The ConceptNet model had more ``non-sequitur" suggestions which may have deterred participants from selecting it.

While there were relatively few added human suggestions, some participants did note that they liked certain features from the AI-generated sets. 
We found that the format of many of the user input game feature suggestions matched the same grammar format we used to extract the training features in subsection \ref{traindat}. Other suggestions provided the reasoning behind why a certain feature could improve the gameplay. This conversational format could be used in the subsystem to provide explainability for the generated features. 

\section{Comparison to ChatGPT}

At the time of writing, transformer models are undergoing rapid development, especially with the development of OpenAI's transformer models ChatGPT\cite{schulman2022chatgpt}. Many arguments could be made in favor of using these larger models over its predecessor GPT-2. However, we would like to note that weights for ChatGPT are (at this time) not publicly accessible. The large number of weights makes it challenging to fine-tune on a particular dataset to be only used for a single function.

To further argue on the usage of GPT-2, we provide excerpts from ChatGPT requesting game features for the same 6 game prompts - prompted by asking ``Can you recommend me 5 different potential game features that are 4-7 words each in bullet point format for the following one sentence game description:" followed by the one-sentence prompt used in the experiments. Table \ref{tab:ChatGPT_out} shows the output responses given by the ChatGPT interaction. While it is impressive how well the model can adapt and stay on theme with the prompt, we find that the features given by ChatGPT are overall subpar in comparison. Many of the features suggested are too generic, repeat features across prompts, or reiterate the description of the game. Very little is offered in terms of specific and innovative game mechanics. While prompt engineering and refinement could be performed to get better results~\cite{reynolds2021prompt}, we plan to continue using the less computationally and financially expensive, GPT-2 model in our system.

\section{Conclusion}

This paper introduces the preliminary work of a conceptual game feature recommendation system and generator. Based on the results of our small user study, and given more time fine-tuning and more feature data, the transformer could compete on the same level as the human-suggested features. Future experiments will look to find an automatic way to select features to show users from the generation model or offer explainability for those generations. Finally, we look to extend this work into a much bigger game design assistance system to work collaboratively with users in a way that is not limited by the genre of the game or the phase of development.

\section*{Acknowledgments}

In accordance with the terms of usage policies, this work is not affiliated with Valve or Launchbox, and the data is only used for academic purposes. We thank the anonymous participants who voluntarily contributed to the user study.

\bibliographystyle{ieeetr}
\bibliography{bibliography}

\end{document}